\documentclass[twocolumn,pra,showpacs,floatfix]{revtex4}
\usepackage{amsmath}
\usepackage{amsfonts}
\usepackage{amsthm}
\usepackage{graphicx}

\newcommand{\ie}{{\textit{i.e.}},\ }
\newcommand{\ket}[1]{\ensuremath{|#1\rangle}}
\newcommand{\bra}[1]{\ensuremath{\langle #1|}}

\newcommand{\identity}{\ensuremath{\openone}}
\newcommand{\nuc}[2]{\mbox{${}^{#1}\rm #2$}}
\newcommand{\units}[2]{\mbox{$#1\,\text{#2}$}}
\newcommand{\para}{\textit{para}}

\newcommand{\pHH}{\para-hydrogen}
\newcommand{\ortho}{\textit{ortho}}
\newcommand{\HH}{\mbox{$\text{H}_2$}}
\newcommand{\hydride}{\mbox{$\text{Ru(H)}_2\text{(CO)}_2\text{(dppe)}$}}
\newcommand{\precursor}{\mbox{$\text{Ru(CO)}_3\text{(dppe)}$}}
\newcommand{\intermediate}{\mbox{$\text{Ru(CO)}_2\text{(dppe)}$}}

\begin{document}
\title{Implementation of NMR quantum computation\\ with \pHH\ derived high purity quantum states}
\author{M.~S. Anwar}
\email{muhammad.anwar@physics.ox.ac.uk} \affiliation{Centre for
Quantum Computation, Clarendon Laboratory, University of Oxford,
Parks Road, OX1 3PU, United Kingdom}

\author{J.~A. Jones}
\email{jonathan.jones@qubit.org} \affiliation{Centre for Quantum
Computation, Clarendon Laboratory, University of Oxford, Parks
Road, OX1 3PU, United Kingdom}

\author{D.~Blazina}
\email{db30@york.ac.uk} \affiliation{Department of Chemistry,
University of York, Heslington, York, YO10 5DD, United Kingdom}
\author{S.~B. Duckett}
\email{sbd3@york.ac.uk} \affiliation{Department of Chemistry,
University of York, Heslington, York, YO10 5DD, United Kingdom}

\author{H.~A. Carteret}
\email{cartereh@iro.umontreal.ca} \affiliation{LITQ, Departement
d'Informatique et Recherche Op\'erationelle, Pavillon
Andr\'e-Aisenstadt, Universit\'e de Montr\'eal, Montr\'eal,
Qu\'ebec H3C 3J7, Canada}

\date{\today}

\begin{abstract}
We demonstrate the first implementation of a quantum algorithm on
a liquid state nuclear magnetic resonance (NMR) quantum computer
using almost pure states.  This was achieved using a two qubit
device where the initial state is an almost pure singlet nuclear
spin state of a pair of \nuc{1}{H} nuclei arising from a chemical
reaction involving \pHH.  We have implemented Deutsch's algorithm
for distinguishing between constant and balanced functions with a
single query.
\end{abstract}
\maketitle

\section{Introduction}
The practical implementation of quantum algorithms has so far been
dominated by liquid state nuclear magnetic resonance (NMR)
\cite{ernst87, levitt, freeman} techniques \cite{cory96, cory97,
jones98a, chuang98, chuang98a, jones98, vandersypen01}. The many
algorithms successfully implemented on liquid state NMR systems
include, most notably, Deutsch's algorithm \cite{jones98a,
chuang98}, Grover's quantum search \cite{chuang98a, jones98}, and
Shor's factoring algorithm \cite{vandersypen01}. This unrivalled
success can be attributed to the high degree of coherent control
over small spin systems and also to the long decoherence times in
comparison with the timescales of the computation. However, one of
the problems haunting the prospects of liquid state NMR as a
practically scalable and ``truly quantum'' implementation, was
that the states encountered in \textit{almost all} of the previous
implementations were highly mixed \cite{braunstein99,warren97}.

The highly mixed states used in NMR raised two major concerns.
First, there was the issue of initializing the spins into a
well-defined state \cite{divincenzo00,jones00}. It was shown that
one could circumvent this problem \cite{cory96,cory97} by
preparing pseudopure states \cite{gershenfeld,knill98,knill00},
which behave \textit{effectively} (up to a certain scaling factor)
as pure states. This approach worked well for small spin systems,
but cannot be extended to larger systems, because of the
exponential loss in signal with increase in the number of spins
\cite{warren97,jones00}.  Secondly, the highly mixed states
encountered in conventional liquid state NMR are separable
\cite{braunstein99}. As entanglement could not exist in these
systems, some authors suggested that NMR quantum computing could
perhaps be explained using classical models \cite{schack}. With
separable states, even the quantum nature of NMR quantum computing
was under question!

Motivated by these problems, we have previously demonstrated
\cite{anwar03} the preparation of an almost pure two-spin state,
which lies above the entanglement threshold \cite{peres96,
horodecki96a}, for the two hydride \nuc{1}{H} nuclei in the
organometallic compound \hydride, where dppe indicates
1,2-bis(diphenyl\-phosphino)ethane. This was achieved using laser
induced addition of pure \pHH\ to \intermediate\ as discussed
below. We now describe an implementation of the Deutsch algorithm
on this spin system, demonstrating that coherent manipulations of
spins in our pure spin system can also be carried out.

Our experiments solve the two-fold problem of initialization and
separability.  We bypass the need for preparing pseudopure states,
as our computation starts off directly with an almost pure Werner
state \cite{werner89}. Furthermore, the initial purity of the
system lies above the entanglement threshold. We do not claim that
our implementation \textit{actually} involves entangled
states---only that entanglement can be distilled from these
states. The use of entanglement proper in the Deutsch's algorithm
has been discussed in more detail elsewhere \cite{collins}. In
short, the current work constitutes the first demonstration of a
quantum algorithm in liquid state NMR using an almost pure initial
state.

The remainder of the paper is organized as follows. In section
\ref{pHH} we describe the use of \pHH\ to generate spin systems in
almost pure initial states, while in section \ref{Deutsch} we give
a brief summary of Deutsch's problem and the quantum algorithm
devised to solve it. Liquid state NMR can be used to implement the
algorithm as with our  two spin pure state; this is discussed in
section \ref{implementation} and the experimental procedures and
results are described in section \ref{experiment}. Finally, we
present a concluding summary in section \ref{conclusions} and also
outline some possible directions for future work.

\section{\pHH}\label{pHH}
The high spin-state purity in our experiment is a result of an
effect called \pHH\ induced polarization (PHIP) \cite{bowers86,
natterer97, duckett99, duckett03}. The existence of the
\textit{para} spin isomer of dihydrogen molecules, \HH, is a
consequence of the Pauli principle and the symmetrization
postulate \cite{messiah}.  These require the overall wavefunction
of the \HH\ molecule to be antisymmetric with respect to
interchange of the fermionic \nuc{1}{H} nuclei.  The translational
and vibrational wavefunctions of \HH\ are always symmetric, as is
the ground state electronic wavefunction, and so overall
antisymmetry is achieved by choosing the rotational and nuclear
spin wavefunctions such that their product is antisymmetric. It
follows that \HH\ molecules in even rotational states
($\text{J}=0,\;2,\;\ldots$) possess an antisymmetric nuclear spin
wave function and correspond to nuclear spin singlets, termed
\para; similarly, \HH\ molecules in odd rotational states
($\text{J}=1,\;3,\;\ldots$) possess a symmetric nuclear spin wave
function and correspond to nuclear spin triplets, termed \ortho.

Interconversion of the two spin-state isomers of hydrogen is
normally forbidden by angular momentum selection rules, but in the
presence of a paramagnetic catalyst the symmetry of the \HH\
system is broken, allowing these rules to be overcome.  Thus if
hydrogen is cooled in the presence of an appropriate catalyst, it
will become enriched in the (low energy) \para\ form.  Upon moving
away from the catalyst interconversion is again suppressed, and so
the \ortho/\para\ ratio in effect \textit{remembers} the
temperature of the last conversion surface encountered.  A
temperature of \units{20}{K} is sufficiently low to cool $99.8\%$
of \HH\ molecules into the $\text{J}=0$ state \cite{duckett99} and
hence produce essentially pure \pHH.

The sole antisymmetric state of two \nuc{1}{H} nuclei is the
singlet state, and so pure \pHH\ will \textit{also} have a pure
nuclear spin singlet state.  This ability to prepare a pure
initial state is obviously attractive for NMR quantum computing
experiments, but the \pHH\ molecule cannot be used directly for
NMR quantum computing, as it is NMR silent due to its high
symmetry. This can be overcome by means of a chemical reaction,
producing a new molecule, in which the two hydrogen atoms can be
made distinct (I and S) and \textit{can} be separately addressed.
In conventional \pHH\ experiments the chemical reaction is slow in
comparison with the frequency difference between the I and S spins
of the reaction product.  This causes the off-diagonal terms in
the density matrix to dephase, producing a separable state.  This
dephasing can be overcome (neglecting relaxation) \cite{Bargon} by
applying an isotropic mixing sequence \cite{MLEV}, but a much
simpler approach is to ensure that the reaction is rapid in
comparison with the dephasing and relaxation timescales.  We
achieve this through addition of \pHH\ to a highly reactive
species, generated by laser flash photolysis of a stable
precursor.

\begin{figure}
\begin{center}
\includegraphics[scale=0.6]{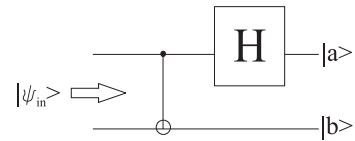}
\caption{Disentangling network, rotating Bell states into
eigenstates. \label{graphicsbell}}
\end{center}
\end{figure}

The state that we actually prepare in our \pHH\ experiments is
observed to be a Werner singlet state \cite{werner89} of the form
\begin{equation}\label{initialstate}
\rho_{init}=\,(1-\varepsilon)\frac{\identity}{4}+\varepsilon\ket{\psi^-}\bra{\psi^-},
\end{equation}
where $\ket{\psi^-}$ is the singlet vector, which can be written
in the computational basis as
\begin{equation}\label{singlet-vector}
\ket{\psi^-}=\,\frac{1}{\sqrt{2}}(\ket{01}-\ket{10}),
\end{equation}
and the polarization $\varepsilon$ determines the purity, with
$\varepsilon=1$ corresponding to the pure state; for our system
\cite{anwar03} $\varepsilon\approx0.92$, and so the state is
almost pure. The singlet state is one of the four Bell states
\cite{nielsen00},
$\ket{\phi^{\pm}}=(\ket{00}\pm\ket{11})/\sqrt{2}$ and
$\ket{\psi^{\pm}}=(\ket{01}\pm\ket{10})/\sqrt{2}$, all of which
can be interconverted using local unitary transformations
\cite{nielsen00,mattle}. The Bell states can also be converted to
basis vectors in the computational basis (the usual starting point
for quantum computations) using a network of the form given in
Figure \ref{graphicsbell}, which is a disentangling network.
The output qubit $\ket{a}$ is determined by the label ($\phi$ or
$\psi$) of the input state $\ket{\psi_{in}}$ and $\ket{b}$ is
determined by the sign ($+$ or $-$); eigenstates can be converted
back to the Bell states, using the reverse of the circuit in Fig.
\ref{graphicsbell}, and the mapping can be compactly expressed as
\begin{equation}
\ket{a}\ket{b}\longleftrightarrow\,\frac{1}{\sqrt{2}}(\ket{0}\ket{b}+(-1)^a\ket{a}\ket{\urcorner
b})),
\end{equation}
where $\ket{\urcorner b}$ is the classical NOT of $\ket{b}$. With
circuits built around the disentangling circuit, we can achieve
conventional two qubit initial states, starting off from
$\ket{\psi_{in}}=\ket{\psi^-}$.

\section{The Deutsch problem\label{Deutsch}}
Deutsch's problem \cite{deutsch92} was the first problem to be
solved by a quantum algorithm. Although its usefulness as a
real-life problem is limited, it is nevertheless a convenient
testing ground for quantum computation. A refined version of
Deutsch's problem \cite{cleve98} is described in the following
way.

Suppose we are given a binary string $\mathbf{s}=\{0,1\}^n$ of
length $n$ and a binary function $f$ acting on $\mathbf{s}$
mapping it onto a single bit, $0$ or $1$, \ie
$f:\{0,1\}^n\rightarrow \{0,1\}$. Furthermore, we are
\textit{promised} that $f$ is either constant or balanced. (A
constant $f$ outputs the \textit{same} value, either $1$ or $0$,
for all $2^n$ possible input strings $\textbf{s}$, whereas a
balanced $f$ outputs $0$ for exactly half of the strings and $1$
for the other half.) The problem is then to find whether $f$ is
constant or balanced using the minimum number of queries. In the
classical setting, answering this question, in the worst case,
would require $(2^n/2)+1$ queries. However, with a quantum
algorithm it is possible to solve this problem in a single query,
giving an exponential improvement over the worst-case classical
version.

The simplest form of Deutsch's problem concerns the case of $n=1$,
with functions mapping a single bit onto a single bit,
$f:\{0,1\}\rightarrow \{0,1\}$. There are four such functions,
their operation being described by the truth table given in Table
\ref{truth-table-f}. Out of these four functions, two are constant
($f_{00}$ and $f_{11}$) and two are balanced ($f_{01}$ and
$f_{10}$). Classically the problem is solved by evaluating
\textit{first} $f(0)$ and \textit{then} $f(1)$ and comparing the
results. Equivalently, $f$ is constant if $f(0)\oplus f(1)=0$ and
balanced if $f(0)\oplus f(1)=1$, where $\oplus$ indicates addition
modulo 2.  Deutsch's algorithm enables us to find this global
property of $f$ in a \textit{single} query.
\begin{table}
\begin{tabular}{ccccc}
\toprule $x$ & $f_{00}(x)$ & $f_{01}(x)$ & $f_{10}(x)$ &
$f_{11}(x)$ \\
 \colrule
$0$ & $0$ & $0$ & $1$ & $1$\\
$1$ & $0$ & $1$ & $0$ & $1$\\
  \botrule
\end{tabular}
\caption{Truth table for the four single-bit functions.}
\label{truth-table-f}
\end{table}

The conventional approach to solve the single-bit problem is to
use two input and two output qubits and then implement $f$ using
carefully selected reversible gates. These gates are represented
by the propagators $U_f$, whose action
on the input qubits can be expressed as 
\begin{equation}
\ket{x}\ket{y}\xrightarrow{U_f}\ket{x}\ket{f(x)\oplus y},
\end{equation}
leaving the first qubit unchanged. If $\ket{y}$ is initialized as
$\ket{0}$, the second qubit holds the value of $f(x)$ after the
computation,
\begin{equation}\label{classical-Deutsch-x0}
\ket{x}\ket{0}\xrightarrow{U_f}\ket{x}\ket{f(x)}.
\end{equation}
There are four propagators, $U_{00}$, $U_{01}$, $U_{10}$ and
$U_{11}$, corresponding to the four different functions. A
suitable choice of propagators is
\begin{align}
U_{00}&=\identity\otimes\identity\label{U00}\\
U_{01}&=\ket{0}\bra{0}\otimes\identity+\ket{1}\bra{1}\otimes\sigma_x\label{U01}\\
U_{10}&=\ket{0}\bra{0}\otimes\sigma_x+\ket{1}\bra{1}\otimes\identity\label{U10}\\
U_{11}&=\identity\otimes\sigma_x\label{U11}
\end{align}
where $\identity$ is the unit matrix of order $2$ and $\sigma_x$
is a Pauli spin matrix \cite{nielsen00}. It is straightforward to
check that with appropriate initial states these propagators
indeed reproduce the transformation given in Equation
\ref{classical-Deutsch-x0}. In the usual language of quantum
networks $U_{00}$ is a do-nothing gate; $U_{01}$ is a
$1$-controlled-NOT gate; $U_{10}$ is a $0$-controlled-NOT gate and
$U_{11}$ is a NOT gate on the second qubit. (Reference
\cite{jones98b} contains a concise description of quantum logic
gates in NMR).

\begin{figure}
\begin{center}
\includegraphics[scale=0.6]{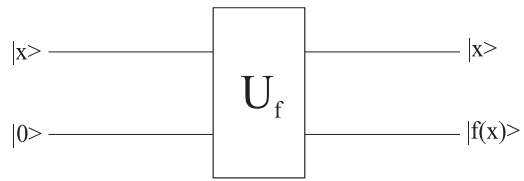}
\caption{Quantum circuit for the classical evaluation of $f(x)$.
\label{graphicsclassical}}
\end{center}
\end{figure}

\begin{figure}
\begin{center}
\includegraphics[scale=0.6]{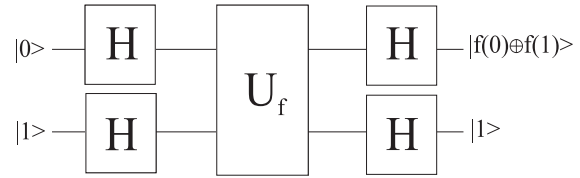}
\caption{Quantum circuit for solving Deutsch's problem. $H$
represents a Hadamard gate. \label{graphicsdeutsch}}
\end{center}
\end{figure}

We can build a quantum circuit from these propagators, which is
depicted in block form in Figure \ref{graphicsclassical}. With
this network, the second qubit evaluates $f(x)$, although the
evaluation is classical and we still need \textit{two} separate
evaluations to determine the nature of $f$, once with $x=0$ and
once with $x=1$. This circuit is, however, useful for illustrative
purposes. For example, using this circuit one could verify the
truth table in Table \ref{truth-table-f}. To solve Deutsch's
problem in a single step, however, we exploit superposition
states, preparing them from the eigenstates using Hadamard gates
\cite{nielsen00}. The corresponding network is given in Figure
\ref{graphicsdeutsch}, and the algorithm proceeds as follows: the
computation starts off in the state $\ket{0}\otimes\ket{1}$, the
Hadamard gates prepare the superposition state, followed by the
unitary transformation $U_f$, and the final Hadamard gates then
convert the superpositions back to eigenstates, with the first
qubit ending in $f(0){\oplus}f(1)$---the solution to the problem!

\section{Implementing the algorithm\label{implementation}}
We have implemented the Deutsch algorithm on an almost pure
($\varepsilon\approx0.92$) two qubit state using an NMR quantum
computer. We also evaluated $f(x)$ classically using the procedure
already outlined. In an NMR quantum computer, the qubits are
spin-$1/2$ atomic nuclei (in this case \nuc{1}{H} nuclei) placed
in a strong magnetic field, and the gates are implemented through
a sequence of radio-frequency (RF) pulses, interspersed with
properly timed delays, during which the spin system evolves under
its own Hamiltonian. Details may be found in reviews of quantum
computing with liquid state NMR, such as \cite{jones98b, cory00,
jones01a, vandersypen04}. Here we use product operator notation
\cite{ernst87, sorensen83, hore} for the states and propagators.
Pulse sequences are written from left to right and we label the
two spins as $I$ and $S$, in keeping with NMR parlance.  The
notation $\theta_{\alpha}$ is used to represent a rotation by
$\theta^{\circ}$ about the $\alpha$ axis.

To implement the classical and quantum versions of the problem
(Figures \ref{graphicsclassical} and \ref{graphicsdeutsch}), the
individual gates have to be built from RF pulses and frame
rotations. For example, the Hadamard gate on a single qubit is a
$180^{\circ}$ rotation about a tilted axis \cite{jones98b} and can
be achieved using the sequence $180_z\,90_{-y}$. Pulse sequences
can also be simplified based on the properties of product
operators: for example, if the Hadamard acts on an eigenstate, the
initial $z$-pulse can be dropped out as these states commute with
$z$-pulses. We are therefore simply left with $90_{-y}$
pseudo-Hadamard operations which can be used as a simpler
replacement for proper Hadamard gates. Likewise, if a computation
ends in a diagonal state, we can shift the $180_z$ pulses to the
end (right) of the sequence, changing, of course, the phase of the
$90^{\circ}$ pulses, and \textit{collapse} the $z$-pulse with the
final state. We are then left with a simple $90_y$ pulse in place
of a strict Hadamard. In the same spirit, the $z$-rotations can be
moved within a sequence, either to the left or to the right, and
can be viewed as rotations of the reference frame, often referred
to as the abstract reference frame \cite{knill00}. We have used
the abstract frames approach together with composite $z$-rotations
\cite{cummins00} to simplify our sequences.

The unitary transformations corresponding to the four possible
$U_f$'s are given in Equations \ref{U00}--\ref{U11}. $U_{00}$ is
an identity gate, and translates to ``do nothing'', while $U_{11}$
is a selective NOT gate on the second qubit, $180S_x$. The
remaining two transformations, $U_{01}$ and $U_{10}$ can be
expressed in product operator notation as, for example,
\begin{equation}
90S_{-y}\thickspace 90I_z \thickspace 90S_z \thickspace(-90)2I_zS_z \thickspace 90S_y
\end{equation}
and
\begin{equation}
90S_{-y}\thickspace 90I_{-z} \thickspace 90S_z \thickspace (90)2I_zS_z\thickspace 90S_y
\end{equation}
respectively.  These transformations involve frame rotations,
evolution under scalar coupling terms ($2I_zS_z$) and soft pulses
\cite{freeman}, selectively exciting the $S$ spin. Selective
pulses were implemented using the Jump and Return pulses described
previously \cite{jones99}, so that our final simplified sequences
comprise only hard pulses and delays. The sequences for the
non-trivial transformations, $U_{01}$, $U_{10}$ and $U_{11}$ are
\begin{align}
P_{01}&=90_{-x}[\tau_1]90_{45}[2\tau_1+\tau_2]180_x[\tau_2]90_{135}[\tau_1]90_{-x}\\
P_{10}&=90_x[\tau_1]90_{45}[2\tau_1+\tau_2]180_x[\tau_2]90_{135}[\tau_1]90_{x}\\
P_{11}&=90_y[2\tau_1]90_{-y}90_x
\end{align}
where $\tau_1=1/4\delta$ and $\tau_2=1/4J$, $\delta$ being the
frequency separation between the spin resonances and $J$ the
spin--spin coupling constant, both measured in Hz.

We have so far overlooked another important issue, the preparation
of initial states. The \pHH\ derived state is an almost pure
singlet, but the algorithms expect the quantum computer to begin
in some eigenstate \ket{ab}, with the values of $a$ and $b$
depending on the algorithm. With circuits built around the
disentangling circuit, Fig. \ref{graphicsbell}, we can achieve the
desired input states, starting off from
$\ket{\psi_{in}}=\ket{\psi^-}$. The simplified pulse sequences to
accomplish this transformation are shown in Table
\ref{bell-to-zeeman}.
\begin{table}
\begin{tabular}{ccccl}
\toprule $a$ & $b$ &&& Pulse sequence \\
 \colrule
$0$ & $0$ &&& \textbf{A}\,$\equiv\,[\tau_1]90_y[\tau_2] 180_x [\tau_2] 180_y [\tau_1] 90_x$ \\
$0$ & $1$ &&& \textbf{B}\,$\equiv\,[\tau_1]90_y[\tau_2] 180_x [\tau_2] 90_y$ \\
$1$ & $0$ &&& \textbf{C}\,$\equiv\,[\tau_1]90_y[\tau_2] 180_x [\tau_2] 90_{-y}$ \\
  \botrule
\end{tabular}
\caption{Pulse sequences for mapping the singlet state
$\ket{\psi^-}$ onto the input states $\ket{ab}$.}
\label{bell-to-zeeman}
\end{table}

Implementation of the classical and Deutsch algorithms involves
preparing the almost pure state $\ket{\psi^-}$, converting it into
the desired initial state using a recipe from Table
\ref{bell-to-zeeman}, and applying the appropriate pulse sequence
for the algorithm. Finally, we need to read out the result at the
end of the computation. At this stage the output states are
diagonal, and so cannot be directly observed, but a simple hard
$90_y$ acquire pulse rotates these states to the measurement
basis: the $\ket{0}$ state gives an NMR signal $I_x$ (positive
absorption) and $\ket{1}$ gives a signal $-I_x$ (negative
absorption). The output states can, therefore, be unambiguously
assigned as $\ket{0}$ or $\ket{1}$ by examining the two
multiplets.

\section{The Experiment\label{experiment}}
The two qubit system for our NMR quantum computer comprises the
two hydride \nuc{1}{H} nuclei in the organometallic compound
\hydride, where dppe indicates 1,2-bis(diphenyl\-phosphino)ethane
and the hydride hydrogen atoms are derived from \pHH.  We prepared
essentially pure \pHH\ at a temperature of \units{18}{K} in the
presence of a charcoal-based catalyst. The gas was introduced into
a \units{5}{mm} NMR tube containing the precursor compound
\precursor, dissolved in d$_6$-benzene. The preparation of the
precursor and finer experimental details are identical to previous
work \cite{anwar03,schott02,godard02}. The NMR tube was then
transferred into a \units{400}{MHz} spectrometer fitted with a
\nuc{1}{H}/\nuc{31}{P} probe modified for \textit{in situ}
photolysis \cite{godard02}. The spectrometer triggered an MPB
Technologies MSX-250 pulsed XeCl excimer laser which fired a
\units{12}{ns} UV pulse of wavelength \units{308}{nm}, irradiating
the active region of the NMR sample and producing the unstable
species \intermediate. This unstable intermediate reacts with
dissolved \pHH\ on the sub-microsecond timescale \cite{cronin95}
leading to the product of interest, \hydride. The two hydrogen
nuclei which inherit the pure singlet spin state from the \pHH\
comprise the two qubits of our quantum computer.

The hydride resonances appear at \units{-7.55}{ppm} (spin $I$) and
\units{-6.32}{ppm} (spin $S$), with a frequency separation of
$\delta=\units{492}{Hz}$. The \nuc{1}{H} transmitter frequency was
placed exactly between the two resonance frequencies, and
couplings to \nuc{31}{P} nuclei were removed by GARP decoupling
\cite{shaka85} applied continuously throughout the experiments.
The $\text{T}_1$ and $\text{T}_2$ constants for the hydride peaks
were measured to be $1.7$ and \units{0.58}{s} respectively while
the hydride $J$ coupling (${}^2J_\text{HH}$) was \units{4.6}{Hz}.
The laser flash acts as an initialisation switch, generating the
pure state $\ket{\psi_{in}}=\ket{\psi^-}$ on demand, which is
subsequently used for the implementation. The pulse sequences for
the classical and quantum evaluations are described below and
summarized in Table \ref{sequences}.
\begin{table}
\begin{tabular}{ccl}
\toprule Experiment && Pulse sequence \\
 \colrule
Classical $f(0)$ && \textbf{A}---\textbf{G}---$U_f$---$90_y$\\
Classical $f(1)$ && \textbf{C}---\textbf{G}---$U_f$---$90_y$\\
Quantum $f(0)\oplus f(1)$ &&\textbf{B}---\textbf{G}---$90_y$---$U_f$\\
  \botrule
\end{tabular}
\caption{Pulse sequences for the Deutsch algorithm. Gradient
fields are represented by \textbf{G}.} \label{sequences}
\end{table}

\begin{figure}
\begin{center}
\includegraphics[scale=0.7]{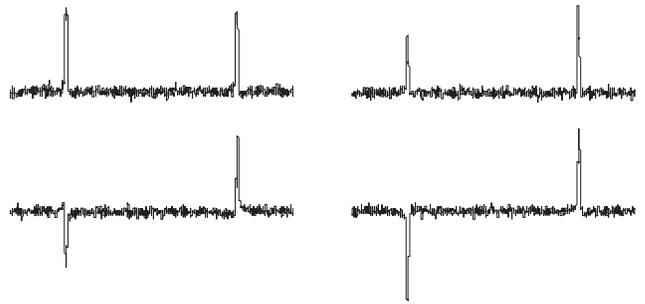}
\caption{Results of the algorithm for the classical determination
of $f(0)$. (a) The result of applying $U_{00}$. The spectrum
corresponds to $I_x+S_x$ and both spins are in positive
absorption. This spectrum can also serve as reference for the
phasing. (b) The result of applying $U_{01}$, corresponding again
to $I_x+S_x$. (c): The result of applying $U_{10}$. The spectrum
now corresponds to $I_x-S_x$, the left spin $I$ is still in
positive absorption and the right spin $S$ is now inverted. (d)
The result of applying $U_{11}$, corresponding to $I_x-S_x$. Spin
$I$ always remains in positive absorption whereas $S$ changes sign
according to $f(0)$.\label{graphics4}}
\end{center}
\end{figure}

Using our system, we were able to analyze $f(0)$ and $f(1)$,
corresponding to the classical evaluation of these functions (see
Figure \ref{graphicsclassical}). For example, for the
determination of $f(0)$ the initial state $\ket{00}$ can be
prepared from the singlet state $\ket{\psi^-}$ using the sequence
\textbf{A} (see Table \ref{bell-to-zeeman}). At this stage, we
clean our state by applying a strong field gradient \cite{freeman,
hore}; this will not affect the component of the state in the
desired eigenstate, but the majority of unwanted terms will be
dephased \cite{jones98}. We then apply the transformation $U_f$
and finally measure with a $90_y$ acquire pulse. The resulting
spectra for the four possible choices of $U_f$ are shown in Figure
\ref{graphics4}. From Table \ref{truth-table-f}, we observe that
$f(0)=0$ for $f_{00}$ and $f_{01}$ and $f(0)=1$ for $f_{10}$ and
$f_{11}$; the result is encoded in the state of spin $S$ (on the
left), whereas spin $I$ (on the right) remains in the state
$\ket{0}$. Thus we should obtain two kinds of spectra,
corresponding to $I_x\pm S_x$, with the $I$ resonance in positive
absorption and the $S$ resonance in positive or negative
absorption, depending on the value of $f(x)$. The two kinds of
spectra are clearly seen (Figure \ref{graphics4}).

\begin{figure}
\begin{center}
\includegraphics[scale=0.7]{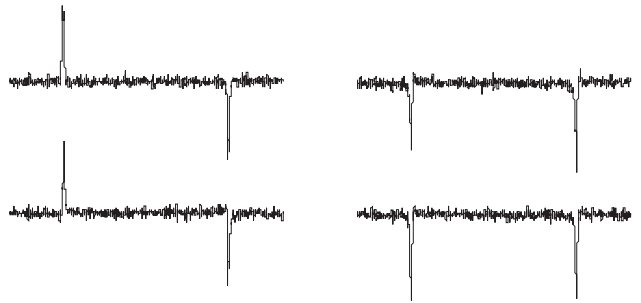}
\caption{Results of the algorithm for the classical determination
of $f(1)$. Labelling and interpretation are the same as in Figure
\ref{graphics4}.  Spin $I$ always remains in negative absorption
whereas $f(1)$ determines the sign of the signal from spin
$S$.\label{graphics5}}
\end{center}
\end{figure}

In the same way we evaluated $f(1)$ by converting the singlet
state into \ket{10} using the sequence \textbf{C} and a strong
gradient field, applying the operation $U_f$ and the acquire
$90_y$ pulse, and observing the result; spectra are shown in
Figure \ref{graphics5}. We again observe two kinds of spectra,
$-I_x\pm S_x$: spin $I$ is always in negative absorption and $S$
changes phase, encoding the value of $f(1)$. Our results are in
complete accord with Table \ref{truth-table-f}.

\begin{figure}
\begin{center}
\includegraphics[scale=0.7]{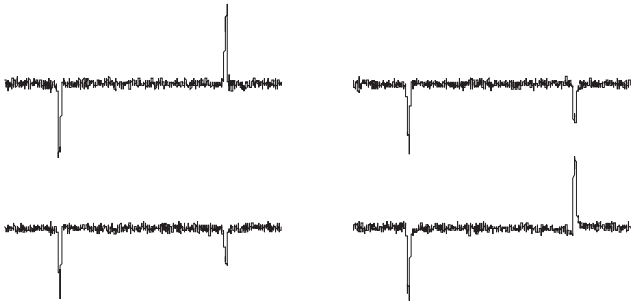}
\caption{Results of the algorithm for the quantum determination of
$f(0)\oplus f(1)$. Labelling and interpretation are the same as in
Figure \ref{graphics4}.  Spin $S$ always remains in negative
absorption while the sign of the signal from spin $I$ encodes the
final result.\label{graphics6}}
\end{center}
\end{figure}

Finally, we implemented the quantum Deutsch algorithm using the
sequence \textbf{B}, followed by the gradient, a pseudo-Hadamard
and the transform $U_f$. An additional simplification occurs in
the quantum version: the final pseudo-Hadamard can be cancelled
with a $90_y$ acquire pulse and the two pulses can therefore be
dropped altogether. The system is, therefore, already in the
measurement basis after the transformation $U_f$ and can be
observed directly. The resulting spectra are shown in Figure
\ref{graphics6}. The observed states after the quantum evaluation
are $I_x-S_x$ and $-I_x-S_x$, for the constant ($f_{00}$,
$f_{11}$) and balanced ($f_{01}$, $f_{10}$) functions
respectively.

Although the overall results are clear, our spectra show several
imperfections.  Each multiplet should in principle show a perfect
absorption lineshape (either positive or negative), and all the
multiplets in all the spectra should ideally have the same
intensity.  In fact dispersive components are clearly visible, and
there are considerable variations in multiplet intensities, both
within spectra and between spectra. These imperfections are most
pronounced in the spectra involving $U_{01}$ and $U_{10}$, which
were obtained with the longest and most complicated pulse
sequences. The distortions can arise from many factors such as
finite coupling evolution during delays, RF inhomogeneities,
miscalibrated pulses and decoherence.

To simulate the effect of decoherence under both $\text{T}_1$ and
$\text{T}_2$ relaxation processes, we employed a model based on
the operator sum representation \cite{vandersypen01,nielsen00} for
phase and generalized amplitude damping. The model is crude and
assumes independent uncorrelated relaxation processes, but is
nonetheless useful in obtaining a rough first estimate of the
effects of decoherence. We observe that the damping processes
should result not only in an overall decrease in the signal
intensity, but for the case of the $U_{01}$ and $U_{10}$
propagators can also lead to intensity imbalances within a
spectrum: for experiments involving these two propagators one of
the spins spends more time in the transverse plane than the other,
and is therefore more prone to phase damping. In the worst cases,
this can cause an overall intensity imbalance of about $2:3$,
similar to the observed ratios. However for a small two qubit
system such as ours, the effects of decoherence are not big enough
to obscure the measurement results.  Comparing the intensities of
the final spectra with the intensity of spectra obtained from the
initial singlet state, we estimate the purity of the final state
to be about $50$ to $60$ percent, indicating that we remain above
the entanglement threshold throughout the implementation.

\section{Conclusions and future work\label{conclusions}}
We have demonstrated an implementation of the Deutsch algorithm to
distinguish between constant and balanced functions using an NMR
quantum computer with an almost pure initial state. We have also
used the same system for classical evaluations of the function
values. Unlike previous implementations built around pseudopure
states, our implementation uses almost pure states which can be
made on demand by laser flash photolysis.  Our earlier work
\cite{anwar03} suggested that we could get around the problem of
low initial polarizations in liquid state NMR by making high
purity states using the \pHH\ approach. The present experiment is
one step forward in showing that the high polarizations can be put
to use in the form of a demonstrable quantum algorithm.

It seems that NMR quantum computing with pure states could
\textit{in principle} be scalable, although more work is needed to
verify the validity (or otherwise) of this claim. We are seeking
to extend this work to larger spin systems and more complex
algorithms using a combination of direct \pHH\ induced
polarization and the transfer of spin-state purity between
different spins.  We are also investigating the possibility of
using \pHH\ techniques to reset qubits midway through a quantum
computation, thus permitting repeated error correction schemes.

\begin{acknowledgments}

We thank the EPSRC for financial support.  MSA thanks the Rhodes
Trust for a Rhodes Scholarship.  HAC thanks MITACS for financial
support.
\end{acknowledgments}

\end{document}